\documentclass[12pt]{article}
\usepackage{latexsym}
\usepackage{amsmath,amssymb}
\usepackage{bm}
\usepackage{graphicx,color}
\usepackage{amssymb}
\usepackage{cite}
\usepackage{amsfonts, amscd, amsthm}
\usepackage[all]{xy}
\usepackage{here}
\usepackage{ulem}
\usepackage{mathabx}
\usepackage[titletoc,title]{appendix}
\usepackage{mathbbol}

\newcommand{\sn}{{\rm sn}}
\newcommand{\cn}{{\rm cn}}
\newcommand{\dn}{{\rm dn}}
\newcommand{\snn}{\widehat{\rm \,sn\,}}
\newcommand{\cnn}{\widehat{\rm cn}}
\newcommand{\dnn}{\widehat{\rm dn}}

\newcommand{\am}{{\rm am}}

\renewcommand{\theequation}{\thesection.\arabic{equation}}
\newcommand{\ee}{{\rm e}}

\newcommand{\dd}{{\rm d}}

\newcommand{\Z}{\mathbb{Z}}

\makeatletter
\@addtoreset{equation}{section}
\renewcommand{\theequation}{\thesection.\@arabic\c@equation}
\makeatother

\makeatletter
\renewcommand\appendix{\par
  \setcounter{section}{0}%
  \setcounter{subsection}{0}%
  \gdef\thesection{Appendix \@Alph\c@section }
  \renewcommand{\theequation}
  {\Alph{section}.\arabic{equation}}
}
\makeatother

\makeatletter
\newcounter{subeqncnt}
\def\thesubeqncnt{\alph{subeqncnt}}
\def\subequations{\begingroup%
\stepcounter{equation}\edef\@tempa{\theequation}%
\let\c@equation\c@subeqncnt\c@subeqncnt\z@
\edef\theequation{\@tempa\noexpand\thesubeqncnt}}

\makeatother

\allowdisplaybreaks

\begin{document}

\titlepage

\title{ The Half-period Addition Formulae 
\\
for Genus Two Hyperelliptic $\wp$ Functions
\\
and
\\
the Sp(4,$\mathbb{R}$) Lie Group Structure} 
\author{Masahito Hayashi\thanks{masahito.hayashi@oit.ac.jp}\\
Osaka Institute of Technology, Osaka 535-8585, Japan\\
Kazuyasu Shigemoto\thanks{shigemot@tezukayama-u.ac.jp} \\
Tezukayama University, Nara 631-8501, Japan\\
Takuya Tsukioka\thanks{tsukioka@bukkyo-u.ac.jp}\\
Bukkyo University, Kyoto 603-8301, Japan\\
}
\date{\empty}


\maketitle
\abstract{ 
In the previous study, by using the two-flows Kowalevski top, we have demonstrated 
that  the genus two hyperelliptic functions 
provide the Sp(4,$\mathbb{R}$)/$\Z_2$  
$\cong$ SO(3,2) Lie algebra structure. 
In this study, by directly using the differential equations of the genus 
two hyperelliptic $\wp$ functions instead of using 
integrable models, 
we demonstrate that the half-period addition 
formula for the genus two hyperelliptic functions provides 
the order two Sp(4,$\mathbb{R}$) Lie group structure.
}
%
\vspace{10mm}
\section{Introduction} 
\setcounter{equation}{0}

We are interested in the mechanism why there are exact solutions, and 
further a series of infinitely many solutions in some cases,  
for some special non-linear differential equations.
Soliton equations are the examples of such equations, hence the various methods
for studying the soliton systems are beneficial for our objective.
Starting from the inverse scattering 
method~\cite{Gardner,Lax,Zakhrov},  the soliton theory has many interesting 
developments,  such as the AKNS formulation~\cite{Ablowitz}, 
geometrical approach~\cite{Bianchi,Hermann,Sasaki,Reyes},
B\"{a}cklund transformation~\cite{Wahlquist,Wadati1,Wadati2}, 
Hirota equation~\cite{Hirota1,Hirota2}, 
Sato theory~\cite{Sato}, 
vertex construction of the soliton solution~\cite{Miwa1,Date1,Jimbo1}, 
and Schwarzian type mKdV/KdV equation~\cite{Weiss}. 

We expect there is a Lie group structure behind some non-linear differential equation, which 
is the reason why such non-linear differential equation has a 
series  of infinite solutions. 
Owing to the addition formula of the Lie group structure, 
there is a series of infinitely many 
solutions. As the representation of the addition formula of the Lie group,
the algebraic  functions such as trigonometric/elliptic/hyperelliptic
functions will emerge for solutions of special partial differential equations.

The AKNS formalism for the Lax pair is 
a  powerful tool to examine the Lie algebra structure 
of the soliton equations of the non-linear integrable models.
In our previous researches, 
we deduced the 
SL(2,$\mathbb{R}$)/$\Z_2$  $\cong$ SO(2,1) $\cong$ SU(1,1)/$\Z_2$ 
$\cong$ Sp(2, $\mathbb{R}$)/$\Z_2$  
Lie algebra structure for the two-dimensional KdV/mKdV/sinh-Gordon models; in addition,
the SO(3,2) $\cong$ Sp(4,$\mathbb{R}$)/$\Z_2$  Lie algebra structure for the
two-flows (two-dimensional) Kowalevski model~
\cite{Hayashi1,Hayashi2,Hayashi3,Hayashi4,Hayashi5,Hayashi6,Hayashi7}.

In this study, in order to examine the Lie group structure, 
instead of the Lie algebra structure, 
we use algebraic functions  such as the elliptic functions and the genus two hyperelliptic functions directly rather than integrable models indirectly.
Especially, by using the half-period addition formula, we have
deduced that there is the order two 
Sp(4,$\mathbb{R}$) Lie group structure for the genus two hyperelliptic $\wp$
functions. 

The paper is organized as follows: 
In section $2$, we demonstrate that the elliptic functions have SO(3) 
Lie group structure via the algebraic addition formula.
In section $3$, we briefly review the genus two Jacobi's inversion problem
to explicitly present the genus two hyperelliptic $\wp$ function.
Then we review the addition formula of the genus two sigma function, which 
is used in the next section.
In section $4$, we first review that the half-period addition formula of the $\wp$ function
gives the order two Sp(2,$\mathbb{R}$) Lie group structure. Next, we demonstrate that 
the half-period addition formulae of the genus two hyperelliptic $\wp$ functions
give the order two Sp(4,$\mathbb{R}$) Lie group structure. 
We devote the final section to the summary and the discussions.   
\section{
The various addition formulae for the elliptic functions} 
\setcounter{equation}{0}
%
We investigate various types of addition formulae for the elliptic functions, 
classified into analytic, algebro-geometric, and algebraic ones.
For the addition formula which includes derivative terms, we define the analytic addition formula.

\subsection{
The various addition formulae for the Weierstrass type 
and Jacobi type elliptic functions}

We first examine the 
SO(3)/SO(2,1) Lie group structure for the elliptic functions.

The Weierstrass' $\wp$ function satisfies the differential equation~\cite{Hancock}
$$
\wp'(u)^2=4\wp(u)^3-g_2 \wp(u)-g_3=4(\wp(u)-e_1)(\wp(u)-e_2)(\wp(u)-e_3).
$$
An analytic addition formula of the Weierstrass $\wp$ function 
is given by
\begin{equation}
\wp(u_1+u_2)=-\wp(u_1)-\wp(u_2)+\frac{1}{4} 
\left( \frac{\wp'(u_2)-\wp'(u_1)}{\wp(u_2)-\wp(u_1)}\right)^2 .
\label{2e1}
\end{equation}
While, an algebro-geometric addition formula is given by
\begin{eqnarray}
&& x_3=-x_1-x_2+\frac{1}{4}\left(\frac{y_2-y_1}{x_2-x_1}\right)^2 ,
\label{2e2}\\
&& y_3=-\left(\frac{y_2-y_1}{x_2-x_1}\right) (x_3-x_1)-y_1 ,
\label{2e3}\\
&&y_i^2=4 x_i^3-g_2 x_i-g_3, \quad (i=1,2,3), 
\label{2e4}
\end{eqnarray}
which constitutes the Mordell-Weil group in number theory.
In addition, there is an algebraic addition formula, which will be discussed
 in the subsequent subsection.

The Jacobi's $\sn$ function satisfies the differential equation~\cite{Hancock}
$$
(\sn'(u))^2=(1-\sn^2(u))(1-k^2 \sn^2(u)),
$$
with $\sn(0)=0$.
Using the $\sn$ function, we define $\cn$ and $\dn$ functions 
via
$$\sn^2(u)+\cn^2(u)=1 \quad\mathrm{and}\quad k^2\ \sn^2(u)+\dn^2(u)=1,$$
with $\cn(0)=\dn(0)=1.$
An analytic addition formula of the Jacobi's elliptic function is given by
\begin{equation}
\sn(u_1+u_2)=\frac{ \sn(u_1) \sn'(u_2)+\sn'(u_1) \sn(u_2)}
{1-k^2 \sn^2(u_1) \sn^2(u_2)} .
\label{2e5}
\end{equation}
While, by using the Abel's addition theorem, the algebro-geometric addition formula is given by\footnote{We must notice 
that the last term of the numerator in  the right-hand side of 
Eq.(\ref{2e7}) is missing in the Baker's textbook~\cite{Baker1}, p.208.
}
\begin{align}
& x_3=-\frac{x_1 y_2+x_2 y_1}{1-k^2 x_1^2 x_2^2} ,
\label{2e6}\\
& y_3=\frac{y_1 y_2 \left(1+k^2 x_1^2 x_2^2\right) -\left(1-k^2\right) x_1 x_2 \left(1-k^2 x_1^2 x_2^2\right)
-2 k^2 x_1 x_2\left(1-x_1^2\right)\left(1-x_2^2\right)}{\left(1-k^2 x_1^2 x_2^2\right)^2}  , 
 \label{2e7}\\
&y_i^2=\left(1-x_i^2\right)\left(1-k^2 x_i^2\right), \quad (i=1,2,3) . 
\label{2e8}
\end{align}
In addition, algebraic addition formulae are given by
\begin{eqnarray}
&& \sn(u_1+u_2)=\frac{ \sn(u_1) \cn(u_2)\dn(u_2)+ \sn(u_2) \cn(u_1)\dn(u_1)}
{1-k^2 \sn^2(u_1) \sn^2(u_2) } ,
\label{2e9}\\
&& \cn(u_1+u_2)=\frac{ \cn(u_1) \cn(u_2)-\sn(u_1) \dn(u_1) \sn(u_2) \dn(u_2)}
{1-k^2 \sn^2(u_1) \sn^2(u_2) } ,
 \label{2e10}\\
&& \dn(u_1+u_2)=\frac{ \dn(u_1) \dn(u_2)-k^2 \sn(u_1) \cn(u_1) \sn(u_2) \cn(u_2)}
{1-k^2 \sn^2(u_1) \sn^2(u_1) } .
\label{2e11}
\end{eqnarray}
These algebraic addition formulae can be rearranged in the 
relation of the SO(3) 
Lie group elements of the form~\cite{Shigemoto1,Shigemoto2}
\begin{equation}
U(u_1) V(u_3) U(u_2)=V(u_2) U(u_3) V(u_1) , \quad
(u_3=u_1+u_2), 
\label{2e12}
\end{equation}
with
\begin{equation}
U(u)=\exp[ i\ \am(u,k)\ J_3 ] , \quad V(u)=\exp[ i\ \am(k u,1/k)\ J_1] , 
\label{2e13}
\end{equation}
where $\am(u,k)$  is the Weierstrass' $\am$ function 
defined by $\sn(u,k)=\sin(\am(u,k))$ and 
$\cn(u,k)=\cos(\am(u,k))$, 
and $J_1$ and $J_3$ imply spin representations of SO(3).  
More explicitly,  by using the spin 1 representation of $J_1$ and $J_3$,
$$
 J_3=\left( \begin{array}{@{\,}ccc@{\,}}
 0 & -i & 0 \\
 i & 0 & 0 \\
 0   &  0  & 0
 \end{array} \right ) , \quad 
J_1=\left( \begin{array}{@{\,}ccc@{\,}}
 0 & 0            & 0           \\
  0 &  0 & -i \\
  0 & i &  0
 \end{array} \right)  , 
$$
we can express the algebraic addition formula in the relation 
of the Lie group elements $U(u)$ and $V(u)$ of the form
\begin{align}
 &
\left( \begin{array}{@{\,}ccc@{\,}}
  \cn(u_1) & \sn(u_1) & 0 \\
 -\sn(u_1) & \cn(u_1) & 0 \\
  0        & 0        & 1
 \end{array} \right )\!\!
\left( \begin{array}{@{\,}ccc@{\,}}
  1 & 0            & 0           \\
  0 &     \dn(u_3) & k\,\sn(u_3) \\
  0 & -k\,\sn(u_3) &    \dn(u_3)
 \end{array} \right) \!\!
 \left( \begin{array}{@{\,}ccc@{\,}}
   \cn(u_2) & \sn(u_2) & 0 \\
  -\sn(u_2) & \cn(u_2) & 0 \\
   0        & 0        & 1
 \end{array} \right ) \notag\\ 
  =&\left( \begin{array}{@{\,}ccc@{}}
  1 & 0            & 0           \\
  0 &     \dn(u_2) & k\,\sn(u_2) \\
  0 & -k\,\sn(u_2) &    \dn(u_2)
 \end{array} \right)\!\!
  \left( \begin{array}{@{\,}ccc@{\,}}
  \cn(u_3) & \sn(u_3) & 0 \\
 -\sn(u_3) & \cn(u_3) & 0 \\
  0        & 0        & 1
 \end{array} \right)\!\!
 \left( \begin{array}{@{\,}ccc@{}}
  1 & 0            & 0           \\
  0 &     \dn(u_1) & k\,\sn(u_1) \\
  0 & -k\,\sn(u_1) &    \dn(u_1)
 \end{array} \right),
\label{2e14}
\end{align}
%
with $u_3=u_1+u_2$. We have used the relations 
$\sn(ku, 1/k)=k\,\sn(u,k)$, $\cn(ku, 1/k)= \dn(u,k)$,  
and $\dn(ku, 1/k)=\cn(u,k)$.
This is the integrability condition, called the Yang-Baxter relation, in the 
two-dimensional integrable statistical model.
%

The elliptic function is formulated with complex numbers, hence we cannot distinguish
between the SO(3) Lie group structure and the SO(2,1) Lie group structure, 
because we can ``analytically continue" from one to another Lie group structure.
For the soliton model, the soliton solution is assumed to be the real number, 
hence the Lie group structure is fixed to be SO(2,1).

\subsection{Algebraic addition formulae for the Weierstrass' $\wp$ function}

In order to obtain an algebraic addition formula for the $\wp$ 
function, we use relations between the $\wp$ function and the Jacobi's elliptic functions in the form~\cite{Hancock}
\begin{equation}
\frac{1}{\sn^2(u)}       =\frac{\wp(z)-e_3}{e_1-e_3}, \quad
\frac{\dn^2(u)}{\sn^2(u)}=\frac{\wp(z)-e_2}{e_1-e_3}, \quad
\frac{\cn^2(u)}{\sn^2(u)}=\frac{\wp(z)-e_1}{e_1-e_3}, 
\label{2e15}
\end{equation}
with $z=u/\sqrt{e_1-e_3}$.  
Noticing that 
\begin{equation}
\displaystyle{\sn(u+3 i K')=\frac{1}{k\ \sn(u)}}, \quad 
\textrm{with}\quad  
\displaystyle{K'=i \int^1_{1/k} \frac{\dd t}{\sqrt{(1-t^2)(1-k^2 t^2)}}},
\label{eq_sn_00}
\end{equation}
$\sn(u)$ and $1/(k\ \sn(u))$ satisfies the same differential equation,
thus they
have similar properties. 
Similarly, we have 
\begin{equation}
\displaystyle{\cn(u+3 i K')=\frac{i\ \dn(u)}{k\ \sn(u)}}, \quad  
\displaystyle{\dn(u+ 3 i K')=\frac{i\ \cn(u)}{\sn(u)}}.
\label{eq_cn} 
\end{equation}
%

Accordingly, we define $\snn(u)$, $\cnn(u)$ 
{\color{red}\sout{,}}
and $\dnn(u)$ functions as
\color{black}
\begin{align}
\snn(u)&=\frac{1}{k\ \sn(u)}        =\frac{\sqrt{\wp(z)-e_3}}{k \sqrt{e_1-e_3}},
\label{2e16}\\
\cnn(u)&=\frac{i\ \dn(u)}{k\ \sn(u)}=\frac{i \sqrt{\wp(z)-e_2}}{k \sqrt{e_1-e_3}},
\label{2e17}\\
\dnn(u)&=\frac{i\ \cn(u)}{\sn(u)}   =\frac{i \sqrt{\wp(z)-e_1}}{\sqrt{e_1-e_3}}  .
\label{2e18}
\end{align}
They satisfy the relations ${\snn}^2(u)+{\cnn}^2(u)=1$ and $k^2 {\snn}^2(u)+{\dnn}^2(u)=1$.

By using the addition formulae of the Jacobi's elliptic functions, we obtain
those of the $\snn(u)$, $\cnn(u)$
and $\dnn(u)$ functions as follows
{\color{red}
\sout{.} :
}
\begin{align}
  \frac{1}{k\ \snn(u_1+u_2)}
&=\frac{ \snn(u_1) \cnn(u_2) \dnn(u_2) + \snn(u_2) \cnn(u_1)\dnn(u_1)}
       {1-k^2 \snn^2(u_1) \snn^2(u_2) } ,
\label{2e19}\\
  \frac{i\ \dnn(u_1+u_2)}{k\  \snn(u_1+u_2)}
&=\frac{ \cnn(u_1) \cnn(u_2) -\snn(u_1) \dnn(u_1) \snn(u_2) \dnn(u_2)}
       {1-k^2 \snn^2(u_1) \snn^2(u_2) } ,
\label{2e20}\\
  \frac{i\ \cnn(u_1+u_2)}{\snn(u_1+u_2)}
&=\frac{\dnn(u_1) \dnn(u_2) - k^2 \snn(u_1) \cnn(u_1) \snn(u_2) \cnn(u_2)}
       {1-k^2 \snn^2(u_1) \snn^2(u_1) } .
\label{2e21}
\end{align}
Eqs.(\ref{2e19})-(\ref{2e21}) imply  the addition formula of  the $\wp(u)$ 
function via Eqs.(\ref{2e16})-(\ref{2e18}).
By using Eqs.(\ref{eq_sn_00})-(\ref{eq_cn})
with the same $K'$ in (\ref{eq_sn_00}), we can prove that 
$\snn$, $\cnn$ and $\dnn$ have the same  
property, hence Eqs.(\ref{2e19})-(\ref{2e21}) are expressed  in the convenient forms  
\begin{align}
  \snn(u_1+u_2+3 i K')
&=\frac{ \snn(u_1) \cnn(u_2) \dnn(u_2) + \snn(u_2) \cnn(u_1)\dnn(u_1)}
       {1-k^2 \snn^2(u_1) \snn^2(u_2) } ,
\label{2e22}\\
  \cnn(u_1+u_2+3 i K')
&=\frac{ \cnn(u_1) \cnn(u_2) - \snn(u_1) \dnn(u_1) \snn(u_2) \dnn(u_2)}
       {1-k^2 \snn^2(u_1) \snn^2(u_2) } ,
\label{2e23}\\
  \dnn(u_1+u_2+3 i K')
&=\frac{\dnn(u_1) \dnn(u_2) - k^2 \snn(u_1) \cnn(u_1) \snn(u_2) \cnn(u_2)}
       {1-k^2 \snn^2(u_1) \snn^2(u_1) }  .
\label{2e24}
\end{align}
We can express Eqs.(\ref{2e22})-(\ref{2e24}) in the relation of the Lie group elements of the form
\begin{align}
 &
\left( \begin{array}{@{\,}ccc@{\,}}
  \cnn(u_1) & \snn(u_1) & 0 \\
 -\snn(u_1) & \cnn(u_1) & 0 \\
  0         & 0         & 1
 \end{array} \right )\!\!
\left( \begin{array}{@{\,}ccc@{}}
  1 & 0             & 0            \\
  0 &     \dnn(u_3) & k\,\snn(u_3) \\
  0 & -k\,\snn(u_3) &    \dnn(u_3)
 \end{array} \right) \!\!
 \left( \begin{array}{@{\,}ccc@{\,}}
  \cnn(u_2) & \snn(u_2) & 0 \\
 -\snn(u_2) & \cnn(u_2) & 0 \\
  0         & 0         & 1
 \end{array} \right ) \notag\\ 
  =&\left( \begin{array}{@{\,}ccc@{}}
  1 & 0             & 0            \\
  0 &     \dnn(u_2) & k\,\snn(u_2) \\
  0 & -k\,\snn(u_2) &    \dnn(u_2)
 \end{array} \right)\!\!
  \left( \begin{array}{@{\,}ccc@{\,}}
  \cnn(u_3) & \snn(u_3) & 0 \\
 -\snn(u_3) & \cnn(u_3) & 0 \\
  0         & 0         & 1
 \end{array} \right)\!\!
 \left( \begin{array}{@{\,}ccc@{}}
  1 & 0             & 0            \\
  0 &     \dnn(u_1) & k\,\snn(u_1) \\
  0 & -k\,\snn(u_1) &    \dnn(u_1)
 \end{array} \right),
\label{2e25}
\end{align}
with $u_3=u_1+u_2+3 i K'$.

\section{The Rosenhain's solution for the genus two Jacobi's inversion 
problem}
\setcounter{equation}{0}
The Weierstrass-Klein type approach to the Jacobi's inversion problem  is quite useful to 
observe the whole structure of the Jacobi's inversion problem.  However, 
it is difficult to obtain explicit expressions of the sigma function for higher 
genus  hyperelliptic $\wp$ functions.

\subsection{The Jacobi's inversion problem for the elliptic function}
It is instructive to examine the Jacobi's inversion problem for the elliptic 
function in order to observe the genus two Jacobi's inversion 
problem.

We adopt the elliptic curve 
$$
y^2=4 x^3-g_2 x -g_3=4(x-e_1)(x-e_2)(x-e_3)
$$
and consider the problem of finding the function $u=u(x)$
\begin{equation}
u=\int_{\infty}^x \frac{ \dd x}
{\sqrt{4(x-e_1)(x-e_2)(x-e_3)}} .
\label{3e1}
\end{equation}
Then, Jacobi's inversion problem of 
obtaining the function $x=x(u)$ is solved by introducing the 
theta function in such a way as expressing $x$ as 
a function of the ratio of the theta functions~\cite{Hancock}, i.e.,  
\begin{equation}
x=x(u)=\wp(u)=e_3+(e_1-e_3) 
\left( \frac{\vartheta\left[\begin{array}{@{\,}c@{\,}} 1\\ 0 \\ \end{array}\right](0)\ 
             \vartheta\left[\begin{array}{@{\,}c@{\,}} 0\\ 1 \\ \end{array}\right](u/2\omega_1)}
            {\vartheta\left[\begin{array}{@{\,}c@{\,}} 0\\ 0 \\ \end{array}\right](0)\ 
             \vartheta\left[\begin{array}{@{\,}c@{\,}} 1\\ 1 \\ \end{array}\right](u/2\omega_1)} \right)^2 , 
\label{3e2}
\end{equation}
with
\begin{equation}
\omega_1=\int_{\infty}^{e_1} \frac{ \dd x}{\sqrt{4(x-e_1)(x-e_2)(x-e_3)}} .
\label{3e3}
\end{equation}
We must notice that the $\wp(u)$ is the quadratic function of 
the ratio of the theta functions instead of the 
linear function. Furthermore, the argument of the 
theta function becomes $u/2\omega_1$ instead of the simple $u$.
By introducing the sigma function as the potential of the $\wp$ function 
in the form $\sigma(u)=\ee^{\eta_1 u^2/2 \omega_1} 
\vartheta\left[\begin{array}{@{\,}c@{\,}} 1\\ 1 \\ \end{array}\right](u/2\omega_1)$,
we can simply express $\wp$ function in the form
\begin{eqnarray}
&&\wp(u)=-\frac{\dd^2}{\dd u^2}  \log \sigma(u)=-\frac{\eta_1}{\omega_1}
-\frac{\dd^2}{\dd u^2} \log \vartheta\left[\begin{array}{@{\,}c@{\,}} 1\\ 1 \\ 
\end{array}\right](u/2\omega_1), 
\label{3e4} 
\end{eqnarray}
where $\eta_1=\zeta(\omega_1)$.
 The role of the factor $\ee^{\eta_1 u^2/2 \omega_1} $ is to shift the constant value
of the $\wp$ function in such a way as $\wp(u)$ has no constant term 
in the Laurent expansion around $u \approx 0$ in the form
$\wp(u)=1/u^2+g_2/20\ u^2+g_3/28\ u^4+\cdots$, which 
is equivalent to set $\lambda_2=0$ in the elliptic curve of the 
form $y^2=4 x^3+\lambda_2 x^2+\lambda_1 x+\lambda_0$.
 
\subsection{The genus two Jacobi's inversion problem}
%
The genus two hyperelliptic $\wp$ functions were given 
by G\"{o}pel~\cite{Gopel1,Gopel2} and  independently by 
Rosenhain~\cite{Rosenhain1,Rosenhain2} via the solution of the  Jacobi's inversion problem . 
However, they are too complicated
to derive the addition formula of the sigma function; which is used in the next section.
Nowadays, G\"{o}pel and Rosenhain's results are little known.
Hence, we sketch 
the Rosenhain's solution for the genus two Jacobi's inversion problem~\cite{Shigemoto3}, 
which provides the explicit 
expressions of $\wp_{22}(u_1,u_2)$ and $\wp_{12}(u_1,u_2)$ by the theta functions.
For the genus two case, we adopt Jacobi's standard form of the hyperelliptic curve
in the form $y^2=x (1-x) (1-k_0^2 x)  (1-k_1^2 x)  (1-k_2^2 x)=f_5(x)$. 
By using three theta function identities, we can consistently parametrize as
\begin{align}
&\left(\frac{\vartheta\left[\begin{array}{@{\,}cc@{\,}} 1 & 0 \\ 1 & 1 \\ \end{array}\right]\!(u_1,u_2)}
       {\vartheta\left[\begin{array}{@{\,}cc@{\,}} 0 & 0 \\ 1 & 1 \\ \end{array}\right]\!(u_1,u_2)}\right)^2
=k_0 k_1 k_2 x_1 x_2 , 
\label{3e5}\\
&\left(\frac{\vartheta\left[\begin{array}{@{\,}cc@{\,}} 1 & 0 \\ 0 & 1 \\ \end{array}\right]\!(u_1,u_2)}
       {\vartheta\left[\begin{array}{@{\,}cc@{\,}} 0 & 0 \\ 1 & 1 \\ \end{array}\right]\!(u_1,u_2)}\right)^2
=-\frac{k_0 k_1 k_2}{k'_0 k'_1 k'_2} (1- x_1)(1- x_2)  , 
\label{3e6}\\
&\left(\frac{\vartheta\left[\begin{array}{@{\,}cc@{\,}} 0 & 1 \\ 0 & 1 \\ \end{array}\right]\!(u_1,u_2)}
       {\vartheta\left[\begin{array}{@{\,}cc@{\,}} 0 & 0 \\ 1 & 1 \\ \end{array}\right]\!(u_1,u_2)}\right)^2
=-\frac{k_1 k_2}{k'_0 k_{01} k_{02}} (1-k_0^2 x_1)(1- k_0^2 x_2) , 
\label{3e7}\\
&\left(\frac{\vartheta\left[\begin{array}{@{\,}cc@{\,}} 0 & 1 \\ 0 & 0 \\ \end{array}\right]\!(u_1,u_2)}
       {\vartheta\left[\begin{array}{@{\,}cc@{\,}} 0 & 0 \\ 1 & 1 \\ \end{array}\right]\!(u_1,u_2)}\right)^2
=\frac{k_0 k_2}{k'_1 k_{01} k_{12}} (1- k_1^2 x_1)(1- k_1^2 x_2) , 
\label{3e8}\\
&\left(\frac{\vartheta\left[\begin{array}{@{\,}cc@{\,}} 0 & 0 \\ 0 & 0 \\ \end{array}\right]\!(u_1,u_2)}
       {\vartheta\left[\begin{array}{@{\,}cc@{\,}} 0 & 0 \\ 1 & 1 \\ \end{array}\right]\!(u_1,u_2)}\right)^2
=\frac{k_0 k_1 }{k'_2 k_{01} k_{12}} (1- k_2^2 x_1)(1- k_2^2 x_2) , 
\label{3e9}
\end{align}
with $k'_0=\sqrt{1-k_0^2}$, $k'_1=\sqrt{1-k_1^2}$, $k'_2=\sqrt{1-k_2^2}$,  
$k_{01}=\sqrt{k_0^2-k_1^2}$, $k_{02}=\sqrt{k_0^2-k_2^2}$, and $k_{12}=\sqrt{k_1^2-k_2^2}$.
Combining any two of these five relations, we obtain ten different expressions 
for $x_1+x_2$ and $- x_1 x_2$. The other ten independent ratios of the theta functions are 
expressed by the symmetric function of $x_1, x_2$ 
in such the form as
\begin{equation}
\left(\frac{\vartheta\left[\begin{array}{@{\,}cc@{\,}} 0 & 0 \\ 0 & 1 \\ \end{array}\right]\!(u_1,u_2)}
     {\vartheta\left[\begin{array}{@{\,}cc@{\,}} 0 & 0 \\ 1 & 1 \\ \end{array}\right]\!(u_1,u_2)}\right)^2
=-\frac{F_{01}(x_1) F_{01}(x_2)}{k'_0 k'_1 k'_2 (x_1-x_2)^2}
\left( \frac{\sqrt{f_5(x_1)}}{F_{01}(x_1)}-\frac{\sqrt{f_5(x_2)}}{F_{01}(x_2)} \right)^2 , 
\label{3e10}
\end{equation}
with $F_{01}(x)=x(1-x)$. 

Next, we differentiate Eqs.(\ref{3e5}) and (\ref{3e6})
and express the result with the theta functions by using the addition 
formulae of the theta functions.  In the expression of that addition formulae,  other 
ratios of the theta functions than those of Eqs.(\ref{3e5})-(\ref{3e9}), i.e., Eq.(\ref{3e10}) etc. 
come out. 
Hence, the function $f_5(x)$ naturally emerges in the Jacobi's inversion problem. 
In order to obtain the standard Jacobi's inversion problem, we can deduce the following equations from Eqs.(3.5) and (3.6) by denoting $U_1=\xi_1u_1+\xi_2u_2$, $U_2=\xi_3u_1+\xi_4u_2$, 
\begin{eqnarray}
\dd U_1=\xi_1 \dd u_1+ \xi_2 \dd u_2= \frac{\dd x_1}{\sqrt{f_5(x_1)}}+\frac{\dd x_2}{\sqrt{f_5(x_2)}} ,
\label{3e11}\\
\dd U_2=\xi_3 \dd u_1+ \xi_4 \dd u_2= \frac{x_1 \dd x_1}{\sqrt{f_5(x_1)}}
+\frac{x_2 \dd x_2}{\sqrt{f_5(x_2)}}  ,
\label{3e12}
\end{eqnarray}
where $\xi_i, (i=1,2,3,4)$ are given by  values of the various
theta functions and their derivatives at $u_1=u_2=0$, which take the rather 
complicated expressions. Then $u_1$ and $u_2$ are expressed as
\begin{equation}
u_1=\eta_1 U_1+ \eta_2 U_2, \qquad 
u_2=\eta_3 U_1+ \eta_4 U_2, 
\label{eq_u1u2}
\end{equation}
with 
\[
\left(\begin{array}{@{\,}cc@{\,}} \eta_1 & \eta_2 \\ \eta_3 & \eta_4 \\ \end{array}\right)
=
\left(\begin{array}{@{\,}cc@{\,}} \xi_1  & \xi_ 2 \\ \xi_3  & \xi_4  \\ \end{array}\right)^{-1}
=
\frac{1}{\xi_1 \xi_4-\xi_2 \xi_3}
\left(\begin{array}{@{\,}rr@{\,}} \xi_4  &-\xi_ 2 \\-\xi_3  & \xi_1  \\ \end{array}\right)
\]

By using Eqs.(\ref{3e5}) and (\ref{3e6}), we obtain  
\begin{align}
&\wp_{22}(U_1,U_2)=x_1+x_2=1
+\frac{1}{k_0 k_1 k_2}
   \left(\frac{\vartheta\left[\begin{array}{@{\,}cc@{\,}} 1 & 0 \\ 1 & 1 \\ \end{array}\right]\!(u_1,u_2)}
              {\vartheta\left[\begin{array}{@{\,}cc@{\,}} 0 & 0 \\ 1 & 1 \\ \end{array}\right]\!(u_1,u_2)}\right)^2\!
+\frac{k'_0 k'_1 k'_2}{k_0 k_1 k_2} 
   \left(\frac{\vartheta\left[\begin{array}{@{\,}cc@{\,}} 1 & 0 \\ 0 & 1 \\ \end{array}\right]\!(u_1,u_2)}
              {\vartheta\left[\begin{array}{@{\,}cc@{\,}} 0 & 0 \\ 1 & 1 \\ \end{array}\right]\!(u_1,u_2)}\right)^2 ,
\label{3e13}\\
&\wp_{12}(U_1,U_2)=- x_1 x_2
=-\frac{1}{k_0 k_1 k_2}
   \left(\frac{\vartheta\left[\begin{array}{@{\,}cc@{\,}} 1 & 0 \\ 1 & 1 \\ \end{array}\right]\!(u_1,u_2)}
              {\vartheta\left[\begin{array}{@{\,}cc@{\,}} 0 & 0 \\ 1 & 1 \\ \end{array}\right]\!(u_1,u_2)}\right)^2 .
\label{3e14}
\end{align}
Substituting the expressions of $u_1$ and $u_2$ in (\ref{eq_u1u2}) 
into the right-hand side of
Eqs.(\ref{3e13}) and  (\ref{3e14}), we obtain the functional expression of 
$\wp_{22}(U_1,U_2)$ and $\wp_{22}(U_1,U_2)$. 
The sigma function $\sigma(u_1,u_2)$ is guaranteed to exist as the potential of the 
$\wp$ function from the integrability conditions. However, it seems difficult to obtain 
an explicit form of the sigma function which is expressed by the theta functions.

For the practical use of the sigma function, it is useful to define the sigma 
function in the Taylor expansion  form in such a way as the 
hyperelliptic $\wp$ functions 
satisfy the differential equations.
Here, we adopt the genus two hyperelliptic curve in the Jacobi's standard form 
$y^2= \lambda_5 x^5+\lambda_4 x^4+\lambda_3 x^3+\lambda_2 x^2+\lambda_1 x+\lambda_0$ with 
$\lambda_5=4$ and $\lambda_0=0$,
because we can easily notice  a dual symmetry in this case.
The differential equations are given by~\cite{Baker3}
\begin{align}
&1)\ \wp_{2222} =6 \wp_{22}^2 +\lambda_4\wp_{22} + \lambda_5\wp_{21} 
+ \frac{1}{8}\lambda_3 \lambda_5,
\label{3e15}\\
&2)\ \wp_{2221} =6 \wp_{22} \wp_{21} +\lambda_4\wp_{21}
- \frac{1}{2} \lambda_5\wp_{11} ,
\label{3e16}\\
&3)\ \wp_{2211} =2 \wp_{22} \wp_{11} +4\wp_{21}^2 +\frac{1}{2}\lambda_3\wp_{21},
\label{3e17}\\
&4)\ \wp_{2111} =6 \wp_{21} \wp_{11}-\frac{1}{2}\lambda_1\wp_{22}+\lambda_2\wp_{21}-\lambda_0 ,
\label{3e18}\\
&5)\ \wp_{1111} =6 \wp_{11}^2 +\lambda_1\wp_{21}+\lambda_2\wp_{11}+\frac{1}{8}\lambda_1 \lambda_3
-\lambda_0 \left(3 \wp_{22}+\frac{1}{2}\lambda_4\right), 
\label{3e19}
\end{align}
where we set $\lambda_0=0$. There is the dual symmetry in the form 
 Eq.(\ref{3e15}) $\leftrightarrow$ Eq.(\ref{3e19}), 
Eq.(\ref{3e16}) $\leftrightarrow$ Eq.(\ref{3e18}), Eq.(\ref{3e17}) $\leftrightarrow$ Eq.(\ref{3e17}) 
under $u_1 \leftrightarrow u_2$,
$\lambda_5 \leftrightarrow \lambda_1$, $\lambda_4 \leftrightarrow \lambda_2$,
$\lambda_3 \leftrightarrow \lambda_3$.

One odd sigma function, which  satisfies five differential equations, is given in 
the form
\begin{eqnarray}
\sigma_1(u_1,u_2)=u_1+\frac{\lambda_2}{24} u_1^3- \frac{\lambda_5}{12} u_2^3 
+\mathcal{O}\left(\{u_1, u_2\}^5\right) .
\label{3e20}
\end{eqnarray}
By using $\wp_{i j}(u_1,u_2)=-\partial_i \partial_j \log \sigma_1(u_1, u_2)$, Baker
obtained the addition formula for one sigma function~\cite{Baker2}
\begin{align}
&\frac{\sigma_1(u_1+v_1,u_2+v_2) \sigma_1(u_1-v_1,u_2-v_2)}
      {\sigma_1(u_1,u_1)^2 \sigma_1(v_1, v_2)^2}
\notag\\
=&\wp_{22}(u_1,u_2) \wp_{12}(v_1,v_2)
 -\wp_{12}(u_1,u_2) \wp_{22}(v_1,v_2)
 -\wp_{11}(u_1,u_2)
 +\wp_{11}(v_1,v_2) .
\label{3e21}
\end{align}
By using the dual symmetry $u_1 \leftrightarrow u_2$, $\lambda_5  \leftrightarrow \lambda_1$,
$\lambda_4  \leftrightarrow \lambda_2$, $\lambda_3 \leftrightarrow \lambda_3$,
we obtain another odd sigma function, which satisfies five differential equations.
This another odd sigma function is given in the form  
\begin{eqnarray}
\sigma_2(u_1,u_2)=u_2+\frac{\lambda_4}{24} u_2^3-\frac{\lambda_1}{12} u_1^3 
+\mathcal{O}\left(\{u_1, u_2\}^5\right)   .
\label{3e22}
\end{eqnarray}
By using $\widehat{\wp}_{ij}(u_1,u_2)=-\partial_i \partial_j \log \sigma_2(u_1,u_2)$,
we obtain the addition formula for another sigma function
\begin{align}
&\frac{\sigma_2(u_1+v_1,u_2+v_2) \sigma_2(u_1-v_1,u_2-v_2)}
      {\sigma_2(u_1,u_2)^2 \sigma_2(v_1,v_2)^2}
\notag\\
&=\widehat{\wp}_{11}(u_1,u_2) \widehat{\wp}_{12}(v_1,v_2) 
 -\widehat{\wp}_{12}(u_1,u_2) \widehat{\wp}_{11}(v_1,v_2)
 -\widehat{\wp}_{22}(u_1,u_2)
 +\widehat{\wp}_{22}(v_1,v_2)  .
\label{3e23}
\end{align} 
Therefore, the addition formula of the sigma function changes depending on what kind of 
sigma function we adopt.

\section{The half-period addition formulae} 
\setcounter{equation}{0}
The half-period addition formula for the elliptic/hyperelliptic functions forms the order two group.
We first examine the half-period addition formula  for the elliptic 
function, which will be instructive to observe the half-period addition formula for the 
genus two hyperelliptic functions.

\subsection{The half-period addition formula for the Weierstrass' $\wp$ function}
 
For the genus one case,  we adopt the Weierstrass elliptic curve of the form 
\[y^2=4 x^3-g_2 x -g_3=4(x-e_1)(x-e_2)(x-e_3).\]
The Jacobi's inversion problem is to obtain $x=\wp(u)$ from  
\begin{equation}
u=\int^x_{\infty} \frac{\dd x}{\sqrt{ 4 (x-e_1) (x-e_2) (x-e_3)}}.
\label{4e1}
\end{equation}
Considering on the Riemann surface, if 
$x$ reaches one of the branch points $e_i\ (i=1, 2, 3)$, $u$ reaches 
the corresponding half-period $\omega_i\ (i=1, 2, 3)$,  
\begin{equation}
\omega_i=\int^{e_i}_{\infty} \frac{\dd x}{\sqrt{ 4 (x-e_1) (x-e_2) (x-e_3)}}, \quad (i=1, 2, 3).
\label{4e2}
\end{equation}
The half-period addition formula of the $\wp$ function is given by 
\begin{equation}
\wp(u+\omega_1)=\frac{e_1 \wp(u)+e_1^2+e_2 e_3}{\wp(u)-e_1}
=\frac{a\wp(u)+b}{c\wp(u)+d}  ,
\label{4e3}
\end{equation}
and that of the cyclic permutation of $\{\omega_1, \omega_2,\omega_3\}$ 
and $\{e_1, e_2, e_3\}$.
We have expressed $a=e_1$, $b=e_1^2+e_2 e_3$, $c=1$, $d=-e_1$ in (\ref{4e3}) and 
observe a matrix defined by    
$$
M=\left( \begin{array}{@{\,}cc@{\,}} a & b \\ c  &  d \end{array}\right)
  =\left( \begin{array}{@{\,}cc@{\,}} a & b \\ c  & -a \end{array}\right)
$$
has SL(2,$\mathbb{R}$) $\cong$ Sp(2,$\mathbb{R}$) 
Lie algebra structure. 
Furthermore, we obtain 
$$
M^2=\left( \begin{array}{@{\,}cc@{\,}}
              a^2+bc & ab +bd\\
              a c+cd & bc +d^2
            \end{array}
     \right)
=(2 e_1^2+e_2 e_3)\ \mathbb{1}, 
$$
which is equivalent to 
\begin{equation}
\wp(u+2 \omega_1)=\frac{(a^2+b c) \wp(u)+(a b+b d)}{(a c + c d)\wp(u)+(b c+d^2)}
=\frac{(2 e_1^2+e_2 e_3) \wp(u)}{(2 e_1^2+e_2 e_3)}=\wp(u).
\label{4e4}
\end{equation}
Hence, the half-period addition formula (\ref{4e3}) provides
the order two SL(2,$\mathbb{R}$) $\cong$ Sp(2,$\mathbb{R}$) 
Lie group structure in addition to the SL(2,$\mathbb{R}$) $\cong$ Sp(2,$\mathbb{R}$) 
Lie algebra structure, which suggests that genus one 
Weierstrass' $\wp$ function has SL(2,$\mathbb{R}$)  
$\cong$ Sp(2,$\mathbb{R}$) Lie group structure in the general case. 
%
%

By applying the half-period transformation twice, we obtain the identity 
transformation. Therefore, the half-period transformation forms the order 
two Lie group transformation.
Thus, we first demonstrate the relation between the Lie algebra element 
and the order two Lie group element 
for the general Sp(2$g$,$\mathbb{R}$)\ ($g$=1, 2, $\cdots$) Lie group.
By using the almost complex structure $J$, which is skew symmetric real matrix with $J^2=-\mathbb{1}$,
the Lie algebra element $A$ and the order two Lie group element $G$  satisfy
\begin{equation}
JA+ A^{T}J=0, \quad G^{T}JG=J,\quad  G^2=\mathbb{1} .
\label{4e5}
\end{equation}
By using $G^2=\mathbb{1}$, we 
obtain $G^{T}J=JG$ from $G^{T}JG=J$. 
For the projective representation of any matrix
$M$,  $\text{(const.)} \times M$ is  equivalent to $M$. 
Thus, in the right-hand side of $G^{T}J=JG$, $JG$ is equivalent to $-JG$, hence we 
obtain the Lie algebra 
relation $G^ {T}J=-JG$, which implies 
that the Lie algebra element $A$ becomes also the order two Lie group 
element $G$.
For the Sp(2,$\mathbb{R}$) case, we adopt 
$J=\left( \begin{array}{@{\,}cc@{\,}} 0 & 1 \\ -1  & 0 \end{array}\right)$ and the Lie group transformation is given by
\begin{equation}
 \left( \begin{array}{@{\,}c@{\,}} x' \\ y' \end{array}\right)
=\left( \begin{array}{@{\,}cc@{\,}} a & b \\ c  & d \end{array}\right)
 \left( \begin{array}{@{\,}c@{\,}}x \\ y \end{array}\right)  ,
\label{4e6}
\end{equation} 
with $G=\left( \begin{array}{@{\,}cc@{\,}} a & c \\ b  & d \end{array}\right)$. 
The projective representation is given by
$\displaystyle{x'/y'=\frac{a x/y + b}{c x/y  +d}}$. For the constant multiplied 
group element $\lambda G
=\left( \begin{array}{@{\,}cc@{\,}} \lambda a & \lambda b \\ \lambda c  
& \lambda d \end{array}\right)$,
the projective representation of the transformation is given by
$\displaystyle{x'/y'=\frac{\lambda a x/y + \lambda b}{\lambda c x/y +\lambda d}
=\frac{a x/y + b}{c x/y +d}}$, i.e.,  
$\lambda G$ is equivalent to $G$ for the projective representation.
The above $M$ satisfies $M^TJ+ JM^T=0$, $M^2={\rm (const.)} \mathbb{1}$. 
This implies that $M$ is not only 
the Sp(2,$\mathbb{R}$) Lie algebra element but also the order 
two Sp(2,$\mathbb{R}$) Lie group element.
   
%
\subsection{The half-period addition formula for the genus two hyperelliptic $\wp$ 
functions}
We adopt the genus two hyperelliptic curve in the form 
\begin{equation}
y^2=f_5(x)=4x^5+\lambda_4 x^4+\lambda_3 x^3+\lambda_2 x^2+\lambda_1 x+\lambda_0
=4(x-e_1)(x-e_2)(x-e_3)(x-e_4)(x-e_5) ,
\label{4e7}
\end{equation}
which gives $\lambda_4=-4(e_1+e_2+e_3+e_4+e_5)$, $\cdots$,  
$\lambda_0=-4 e_1 e_2 e_3 e_4 e_5$. In the Riemann surface,  there are 
six branching points
$e_1,e_2,e_3,e_4,e_5$ and $e_6=\infty$. The cuts are drawn from $e_{2i-1}$ to $e_{2i}$ 
$(i=1,2,3)$.
The Jacobi's inversion problem is given by 
\begin{eqnarray}
u_1=\int_{\infty}^{x_1} \frac{\dd t}{\sqrt{f_5(t)}}
+\int_{\infty}^{x_2} \frac{\dd t}{\sqrt{f_5(t)}},
\quad
u_2=\int_{\infty}^{x_1} \frac{ t \dd t}{\sqrt{f_5(t)}}
+\int_{\infty}^{x_2} \frac{t \dd t }{\sqrt{f_5(t)}},
\label{4e8}
\end{eqnarray}
and the genus two hyperelliptic $\wp$ functions are given by
\begin{equation}
\wp_{22}(u)=x_1+x_2, 
\quad
\wp_{21}(u)=-x_1 x_2, 
\quad
\wp_{11}(u)=\frac{F(x_1,x_2)-2y_1 y_2}{4 (x_1-x_2)^2},
\label{4e9}
\end{equation}
with
\[
  F(x_1,x_2)
  =2\lambda_0+\lambda_1(x_1+x_2)+2\lambda_2x_1x_2+\lambda_3x_1x_2(x_1+x_2)+2\lambda_4(x_1x_2)^2+4(x_1x_2)^2(x_1+x_2) .
\]
By setting $x_1=e_i,\,x_2=e_j$, the half-period is given by 
$\Omega=(\omega_1,\omega_2)$ for $u=(u_1,u_2)$ in the form
\begin{eqnarray}
\omega_1=\int_{\infty}^{e_i} \frac{\dd t}{\sqrt{f_5(t)}}
+\int_{\infty}^{e_j} \frac{\dd t}{\sqrt{f_5(t)}},
\quad
\omega_2=\int_{\infty}^{e_i} \frac{ t \dd t}{\sqrt{f_5(t)}}
+\int_{\infty}^{e_j} \frac{ t \dd t}{\sqrt{f_5(t)}},
\label{4e10}
\end{eqnarray}
 which provides $\wp_{22}(\Omega)=e_i+e_j$, $\wp_{22}(\Omega)=-e_i e_j$,
$\wp_{11}(\Omega)=F(e_i,e_j)/4(e_i-e_j)^2$, where we use $y_1(\Omega)=0$ 
and $y_2(\Omega)=0$ because $x_1(\Omega)=e_i$ and $x_2(\Omega)= e_j $.

In order to obtain the half-period  addition formula for the 
hyperelliptic $\wp$ functions, we use the addition formula of 
the sigma function Eq.(\ref{3e21}) of the form 
\begin{equation}
\frac{\sigma(u+v) \sigma(u-v)}{\sigma(u)^2 \sigma(v)^2}=\wp_{22}(u) \wp_{21}(v) -\wp_{21}(u) \wp_{22}(v)
-\wp_{11}(u)+\wp_{11}(v)  .
\label{4e11}
\end{equation}
Next, we set $v=\Omega=(\text{half\ period})$, hence we have 
\begin{align}
\frac{\sigma(u+\Omega)^2}{\sigma(\Omega)^2 \sigma(u)^2}
&=\wp_{21}(\Omega) \wp_{22}(u) -\wp_{22}(\Omega)\wp_{21}(u)
-\wp_{11}(u)+\wp_{11}(\Omega)
\notag\\
&=d_1\ \wp_{22}(u) +d_2\ \wp_{21}(u) +d_3\ \wp_{11}(u)+d_4  ,
\label{4e12}
\end{align}
where $d_1=\wp_{21}(\Omega)$,  $d_2=-\wp_{22}(\Omega)$, $d_3=-1$, $d_4=\wp_{11}(\Omega)$.
Considering the logarithm of Eq.(\ref{4e12}) and differentiating twice, 
we  obtain 
\begin{align}
\wp_{i j}(u+\Omega)
&=\wp_{i j}(u)-\frac{1}{2}\frac{Q(u,\Omega) \partial_i \partial_j Q(u,\Omega)
-(\partial_i Q(u,\Omega) ) (\partial_j Q(u,\Omega) )}{ Q(u,\Omega)^2}
\notag\\
&=\frac{2 Q(u,\Omega)^2 \wp_{i j}(u)-Q(u,\Omega) \partial_i \partial_j Q(u,\Omega)
+(\partial_i Q(u,\Omega) ) (\partial_j Q(u,\Omega)) }{2 Q(u,\Omega)^2},   
\label{4e13}
\end{align}
with
\[
  Q(u,\Omega)=d_1\ \wp_{22}(u) +d_2\ \wp_{21}(u) +d_3\ \wp_{11}(u)+d_4 ,
\]
by using $\wp_{i j}(u_1,u_2)=-\partial_i \partial_j \sigma(u_1, u_2)$.
Using  Eqs.(\ref{3e15})-(\ref{3e19}) and 
Eqs.(\ref{A1})-(\ref{A10}) in
 the Appendix A, the numerator and the denominator are expressed by the polynomial 
of various $\wp_{i j}\mbox{'s}$.  
In the numerator of the right-hand side of Eq.(\ref{4e13}), we have 
the third-degree terms of
$\wp_{i j}\mbox{'s}$ in general, yet the  third-degree terms automatically cancel. 
Therefore, the numerator starts  from the second-degree terms of $\wp_{i j}\mbox{'s}$. 
Furthermore, as it is surprisingly enough, 
the numerator has the factor $Q(u,\Omega)$. Thus, both the numerator and the denominator 
stars from the first degree
terms of $\wp_{i j}\mbox{'s}$.

Hence, the addition formulae for half-period are given by Baker~\cite{Baker2} and 
Buchstaber {\it et al.}~\cite{Buchstaber1} in the form
\begin{eqnarray}
\wp_{22}(u+\Omega )=\frac{a_1 \wp_{22}(u) 
+a_2 \wp_{21}(u) +a_3 \wp_{11}(u)+a_4}
{d_1 \wp_{22}(u) +d_2 \wp_{21}(u) +d_3 \wp_{11}(u)+d_4}  ,
\label{4e14}\\
\wp_{21}(u+\Omega)=\frac{b_1 \wp_{22}(u) 
+b_2 \wp_{21}(u) +b_3 \wp_{11}(u)+b_4}
{d_1 \wp_{22}(u) +d_2 \wp_{21}(u) +d_3 \wp_{11}(u)+d_4}  ,
\label{4e15}\\
\wp_{11}(u+\Omega)=\frac{c_1 \wp_{22}(u) 
+c_2 \wp_{21}(u) +c_3 \wp_{11}(u)+c_4}
{d_1 \wp_{22}(u) +d_2 \wp_{21}(u) +d_3 \wp_{11}(u)+d_4}  .
\label{4e16}
\end{eqnarray}
There are two types of the half-periods. Type I is given by setting
 $x_1(\Omega)=e_i,\,x_2(\Omega)=e_j,\,(i \ne j,\,1\le i, j \le 5)$.  
Type II is given by setting 
$x_1(\Omega)= e_i,\,x_2(\Omega)= e_6=\infty,\,(1\le i \le 5)$.

For the type I half-period addition formula, we consider the following 
example of $e_i=e_1,\,e_j=e_2$,
and use the  expression of  Buchstaber {\it et al.}'s paper. In this case, we have 
the expression $\wp_{22}(\Omega_{{\rm I}})=e_1+e_2$, 
$\wp_{22}(\Omega_{{\rm I}})=-e_1 e_2$,
$\wp_{11}(\Omega_{{\rm I}})=F(e_1,e_2)/4(e_1-e_2)^2=e_1 e_2(e_3+e_4+e_5)+e_3 e_4 e_5$, which provides  
\begin{eqnarray}
&&G_{{\rm I}}=\left( \begin{array}{@{\,}cccc@{\,}} 
a_{1} & a_{2} & a_{3} & a_{4}\\ 
b_{1} & b_{2} & b_{3} & b_{4}\\ 
c_{1} & c_{2} & c_{3} & c_{4}\\ 
d_{1} & d_{2} & d_{3} & d_{4}\\ 
\end{array}\right)
=\left( \begin{array}{@{\,}cccc@{\,}} 
 a_{1} &  a_{2} &  a_{3} & a_{4}\\ 
 b_{1} & -a_{1} &  b_{3} & b_{4}\\ 
 b_{4} & -a_{4} &  c_{3} & c_{4}\\ 
-b_{3} &  a_{3} & -1     &-c_{3}\\ 
\end{array}\right)  .
\label{4e17}
\end{eqnarray}
One of the examples is given by 
\begin{alignat*}{4}
a_1&=S_3-S_1 s_2,     & \ a_2&=-s_2-S_1 s_1+S_2,& \ a_3&=-s_1,& \ a_4&=-s_2^2+S_2 s_2+S_1 s_2 s_1, \\
b_1&=-S_3 s_1+S_2s_2, &   b_2&=-a_1,            &   b_3&=s_2, &   b_4&=-2 S_3 s_2 +S_3 s_1^2-S_2s_2 s_1, \\
c_1&=b_4,             &   c_2&=-a_4,            &   c_3&= -S_3-S_1 s_2 ,\\[-10pt]
\intertext{\hspace{2mm}$  c_4=-S_2^2 s_2+4 S_3 S_1 s_2+S_2 S_1 s_2 s_1+S_3 S_2 s_1+S_2 s_2^2-S_3 S_1 s_1^2 -S_3 s_2 s_1 ,$}\\[-25pt]
d_1&=-b_3,            &   d_2&=a_3,             &   d_3&=-1,  &   d_4&= -c_{3}  ,
\end{alignat*}
with $s_1=e_1+e_2$, $s_2=e_1e_2$, $S_1=e_3+e_4+e_5$,
$S_2=e_3 e_4+e_4 e_5+e_5 e_3$, $S_3=e_3 e_4 e_5$.
All type I half-periods are given by arranging 
$\{e_1,e_2, e_3, e_4,e_5\}$ into two 
sets $\{e_i, e_j\} \cup \{e_p, e_q, e_r\}$. 
We can verify $G_{{\rm I}}^2=\text{(const.)} \mathbb{1}$
for all type I half-periods.

For the type II half-period addition formula, we consider the following example 
of $e_i=e_1, e_j=e_6=\infty$
and we use the  expression of  Buchstaber {\it et al.}'s paper.
We take the most singular term and the limit $e_6 \rightarrow \infty$ at the end.
Thus, we have the expression  
$\wp_{22}(\Omega_{{\rm II}})=e_6$, $\wp_{22}(\Omega_{{\rm II}})=-e_1 e_6$,
$\wp_{11}(\Omega_{{\rm II}})=e_1^2 e_6$, which provides
\begin{eqnarray}
&&G_{{\rm II}}=\left( \begin{array}{@{\,}cccc@{\,}} 
\hat{a}_{1} & \hat{a}_{2} & \hat{a}_{3} & \hat{a}_{4}\\ 
\hat{b}_{1} & \hat{b}_{2} & \hat{b}_{3} & \hat{b}_{4}\\ 
\hat{c}_{1} & \hat{c}_{2} & \hat{c}_{3} & \hat{c}_{4}\\ 
\hat{d}_{1} & \hat{d}_{2} & \hat{d}_{3} & \hat{d}_{4}\\ 
\end{array}\right)
=\left( \begin{array}{@{\,}cccc@{\,}} 
 \hat{a}_{1} & 0           &  1           &  \hat{a}_{4}\\ 
 0           & \hat{a}_{1} &  \hat{b}_{3} &  \hat{b}_{4}\\ 
-\hat{b}_{4} & \hat{a}_{4} & -\hat{a}_{1} & 0           \\ 
 \hat{b}_{3} & -1          &  0           & -\hat{a}_{1}\\ 
\end{array}\right)  .
\label{4e18}
\end{eqnarray}
One of the examples is given by 
\begin{alignat*}{4}
\hat{a}_1&=-e_1^2,    &\quad \hat{a}_2&=0,        &\quad \hat{a}_3&=1,         &\quad \hat{a}_4&=e_1^2 T_1-e_1 T_2, \\
\hat{b}_1&=0,         &\quad \hat{b}_2&=\hat{a}_1,&\quad \hat{b}_3&=-e_1,      &\quad \hat{b}_4&=e_1 T_3-T_4, \\
\hat{c}_1&=-\hat{b}_4,&\quad \hat{c}_2&=\hat{a}_4,&\quad \hat{c}_3&=-\hat{a}_1,&\quad \hat{c}_4&=0, \\
\hat{d}_1&= \hat{b}_3,&\quad \hat{d}_2&=-1,       &\quad \hat{d}_3&=0,         &\quad \hat{d}_4&=-\hat{a}_1, 
\end{alignat*}
with $T_1=e_2+e_3+e_4+e_5$, $T_2=e_2 e_3+e_2 e_4+e_2 e_5+e_3 e_4+e_3 e_5+e_4 e_5$,
$T_3=e_2 e_3 e_4+e_2 e_3 e_5+e_2 e_4 e_5+e_3 e_4 e_5$,
$T_4=e_2 e_3 e_4 e_5$.
All type II half-periods are given by arranging 
$\{e_1,e_2, e_3, e_4,e_5\}$ into two 
sets $\{e_i\} \cup \{e_k, e_p, e_q, e_r\}$. 
We can verify $G_{{\rm II}}^2=\text{(const.)} \mathbb{1}$
for all type II half-periods.
%

%

For Sp(4,$\mathbb{R})$ case, we adopt the representation of almost complex structure $J$ with 
$J^2=-\mathbb{1}$ in the form
~\footnote{
Depending on the ordering of the elements of the vector  
$(\wp_{22}, \wp_{21}, \wp_{11},1)$, the representation of $G_{{\rm I}}$, $G_{{\rm II}}$ and $J$ changes. 
We adopt Baker's ordering~\cite{Baker2}. The existence of the Sp(4,$\mathbb{R}$) 
Lie group structure is independent of such ordering. }
\begin{equation}
J= \left( \begin{array}{@{\,}cccc@{\,}} 0 & -1 & 0 & 0\\ 1 & 0 & 0 & 0\\  0 & 0 & 0 & 1\\ 0 & 0 & -1 & 0\\
\end{array}\right)  .
\label{4e19}
\end{equation}
The bases of the Sp(4,$\mathbb{R}$) Lie algebra,  which satisfies $J A+A^T J=0$,
is given by 
\begin{alignat*}{4}
I_1&=
\begin{pmatrix}
0&1&0&0\\
0&0&0&0\\
0&0&0&0\\
0&0&0&0
\end{pmatrix},
&\
I_2&=
\begin{pmatrix}
0&0&0&0\\
1&0&0&0\\
0&0&0&0\\
0&0&0&0
\end{pmatrix},
&\ 
I_3&=
\begin{pmatrix}
1& 0&0&0\\
0&-1&0&0\\
0& 0&0&0\\
0& 0&0&0
\end{pmatrix},
&\ \ 
I_4&=
\begin{pmatrix}
0&0&1&0\\
0&0&0&0\\
0&0&0&0\\
0&1&0&0
\end{pmatrix},
\\
I_5&=
\begin{pmatrix}
0&0&0&0\\
0&0&0&1\\
1&0&0&0\\
0&0&0&0
\end{pmatrix},
&\
I_6&=
\begin{pmatrix}
0& 0&0&1\\
0& 0&0&0\\
0&-1&0&0\\
0& 0&0&0
\end{pmatrix},
&\
I_7&=
\begin{pmatrix}
 0&0&0&0\\
 0&0&1&0\\
 0&0&0&0\\
-1&0&0&0
\end{pmatrix},
&\
I_8&=
\begin{pmatrix}
0&0&0&0\\
0&0&0&0\\
0&0&0&1\\
0&0&0&0
\end{pmatrix},
\\
I_9&=
\begin{pmatrix}
0&0&0&0\\
0&0&0&0\\
0&0&0&0\\
0&0&1&0
\end{pmatrix},
&\
I_{10}&=
\begin{pmatrix}
0&0&0& 0\\
0&0&0& 0\\
0&0&1& 0\\
0&0&0&-1
\end{pmatrix}.
\end{alignat*}
%
We have verified that $G_{\rm I}$ and $G_{\rm II}$ satisfies $JG_{\rm I}+G_{\rm I}^{T}J=0$, 
and $JG_{\rm II}-G_{\rm II}^{T}J=0$.  
We observe that Eq.(\ref{4e17}) is the element 
of the ordinary Sp(4,$\mathbb{R}$) Lie algebra; yet 
 Eq.(\ref{4e18}) is not that of the ordinary Sp(4,$\mathbb{R}$) Lie algebra.
However, the transformation of the half-period addition formula
for the hyperelliptic $\wp$ functions provides the projective representation.
Hence, in the projective representation, $G_{\rm I}$ and $G_{\rm II}$ are not only 
the elements of  the Sp(4,$\mathbb{R}$)  Lie algebra
but also the elements of the order two Sp(4,$\mathbb{R}$) Lie group.
This suggests that the genus two hyperelliptic functions have the general continuous
Sp(4,$\mathbb{R}$) Lie group structure.

\section{Summary and Discussions} 
\setcounter{equation}{0}

First, we have examined various types of addition formulae for the elliptic functions. 
The algebraic addition formula can be rearranged into the relation of the Lie group elements, which is called the 
Yang-Baxter's integrable condition.
Second, we have 
reviewed the Rosehnain's approach to the genus two Jacobi's inversion problem.
It is difficult to express the explicit form of the sigma function for the genus two case,
thus we use the Taylor expansion form for the sigma function. We pointed out that addition formula of the 
sigma function depends on what kind of sigma function we adopt.
Finally, we have obtained the order two addition formula of the genus two 
hyperelliptic $\wp$ functions by using the addition formula of a sigma function.

In the previous study, via the two flows Kowalevski top, we had demonstrated  
that the genus two hyperelliptic functions provide the 
Sp(4,$\mathbb{R}$)/$\Z$ $\cong$ SO(3,2) Lie algebra structure.
In this study, by directly using the differential equations of the genus 
two hyperelliptic $\wp$ functions, we have demonstrated that the half-period addition 
formula for the genus two hyperelliptic $\wp$ functions provides the order two 
Sp(4,$\mathbb{R}$) Lie group structure.
This suggests that the genus two 
hyperelliptic $\wp$ functions have the general continuous Sp(4,$\mathbb{R}$) 
Lie group structure.

\begin{appendices}
\setcounter{equation}{0}
\section{The $\wp_{ijk} \wp_{lmn}$ type differential equations 
for the genus two hyperelliptic equations} 
The differential equations for the genus two hyperelliptic $\wp$ functions, which corresponds 
to $\wp'^2=4\wp^3-g_2 \wp -g_3$ in the genus one elliptic $\wp$ function, are 
given by~\cite{Buchstaber1}~\footnote{
In the Buchstaber {\it et al.}'s paper,  the last term of equation Eq.(A.4) 
is given by $-\lambda_0 \lambda_3 \lambda_4/4$, 
but this contains a typographical error and is correctly 
given by $-\lambda_0 \lambda_2 \lambda_4/4.$}
\begin{align}
&1)\ &\wp_{222}^2 =&4 \wp_{22}^3 +\lambda_4 \wp_{22}^2
+4 \wp_{22}\wp_{21}+\lambda_3\wp_{22}
+4 \wp_{11} +\lambda_2,
\label{A1}\\
&2)\ &\wp_{221}^2 =&4 \wp_{22}\wp_{21}^2 +\lambda_4 \wp_{21}^2
-4 \wp_{21}\wp_{11}+\lambda_0,
\label{A2}\\
&3)\ &\wp_{211}^2 =&4 \wp_{21}^2\wp_{11} +\lambda_0 \wp_{22}^2
-\lambda_1 \wp_{22}\wp_{21}+\lambda_2\wp_{21}^2,
\label{A3}\\
&4)\ &\wp_{111}^2 =&4 \wp_{11}^3 +\lambda_0 \wp_{21}^2
-4\lambda_0 \wp_{22}\wp_{11}+\lambda_1 \wp_{21}\wp_{11}+\lambda_2\wp_{11}^2
+\left(\frac{1}{4} \lambda_1^2-\lambda_0 \lambda_2\right) \wp_{22}
\notag\\
&& &+\frac{1}{2} \lambda_0 \lambda_3\wp_{21} 
+\left(\frac{1}{4} \lambda_1 \lambda_3-\lambda_0 \lambda_4\right) \wp_{11}
+\frac{1}{16}\left(\lambda_1^2\lambda_4+\lambda_0 \lambda_3^2-4\lambda_0 \lambda_2 \lambda_4\right),
\label{A4}\\
&5)\ &\wp_{222}\wp_{221} =& 4 \wp_{22}^2\wp_{21} +\lambda_4 \wp_{22}\wp_{21}
-2 \wp_{22}\wp_{11}+2 \wp_{21}^2+\frac{1}{2}\lambda_3 \wp_{21}+ \frac{1}{2}\lambda_1,
\label{A5}\\
&6)\ &\wp_{222}\wp_{211} =& 2 \wp_{22}^2\wp_{11} +2 \wp_{22}\wp_{21}^2
+\frac{1}{2}\lambda_3 \wp_{22}\wp_{21}+4 \wp_{21}\wp_{11}-\frac{1}{2}\lambda_1 \wp_{22}
+\lambda_2 \wp_{21}, 
\label{A6}\\
&7)\ &\wp_{222}\wp_{111} =& 6\wp_{22}\wp_{21} \wp_{11}-2 \wp_{21}^3-\lambda_1 \wp_{22}^2
+2 \lambda_2\wp_{22}\wp_{21}-\frac{1}{2}\lambda_3\wp_{22}\wp_{11}
-\lambda_3 \wp_{21}^2
\notag\\
&&& +2 \lambda_4 \wp_{21} \wp_{11}-4\wp_{11}^2-\frac{1}{4} \lambda_1 \lambda_4 \wp_{22}
+\frac{1}{8}\left(-4\lambda_1+4\lambda_2 \lambda_4-\lambda_3^2\right) \wp_{21}
\notag\\
&&&
-\lambda_2 \wp_{11}
-\frac{1}{8}\lambda_1 \lambda_3,
\label{A7}\\
&8)\ &\wp_{221}\wp_{211} =& 2 \wp_{22} \wp_{21}\wp_{11} +2 \wp_{21}^3
+\frac{1}{2}\lambda_3 \wp_{21}^2 -\lambda_0 \wp_{22}+\frac{1}{2}\lambda_1 \wp_{21},
\label{A8}\\
&9)\ &\wp_{221}\wp_{111} =& 2\wp_{22}\wp_{11}^2+2\wp_{21}^2\wp_{11}
-2\lambda_0 \wp_{22}^2+\lambda_1 \wp_{22} \wp_{21}
+\frac{1}{2}\lambda_3\wp_{21}\wp_{11}
\notag\\
&&&-\frac{1}{2} \lambda_0 \lambda_4 \wp_{22}+\frac{1}{4}\left(\lambda_1\lambda_4-4 \lambda_0\right)\wp_{21}
-\frac{1}{2}\lambda_1 \wp_{11}-\frac{1}{4} \lambda_0 \lambda_3,
\label{A9}\\
&10)\ &\wp_{211}\wp_{111} =& 4\wp_{21}\wp_{11}^2-\lambda_0 \wp_{22} \wp_{21}
-\frac{1}{2}\lambda_1 \wp_{22} \wp_{11}+\frac{1}{2}\lambda_1 \wp_{21}^2
+\lambda_2 \wp_{21} \wp_{11}
\notag\\
&&&-\frac{1}{4} \lambda_0 \lambda_3 \wp_{22}
+\frac{1}{8}\lambda_1 \lambda_3 \wp_{21}
-2 \lambda_0 \wp_{11}+\frac{1}{8}\left(-4\lambda_0 \lambda_2+\lambda_1^2\right) .
\label{A10}
\end{align}
\end{appendices}


\end{document}